\documentstyle[preprint,aps]{revtex}
\begin{document}
\preprint{\vbox{\hbox{OITS-633}}}
\draft
\title{A New Method Of 
Distinguishing Models For The High-$Q^2$ Events
At HERA}
\author{Zhen Cao$^1$, Xiao-Gang He$^2$ and Bruce McKellar$^2$}
\address{$^1$Institute of Theoretical Science, University of Oregon,
Eugene, OR 97403-5203, USA\\
$^2$School of Physics,
University of Melbourne,
Parkville, Vic. 3052, Australia}
\date{July, 1997}
\maketitle
\begin{abstract}
Many explanations for the excess high-$Q^2$ $e^+ p \rightarrow e^+ X$ 
events from H1 and ZEUS at HERA have been proposed each with criticisms.
 We propose a new method to  
distinguish different models by looking at a new distribution
which is insensitive to parton distribution function,
but sensitive to new physics.
\end{abstract}

The H1 and ZEUS experiments at HERA have observed \cite{1} 
an excess of events in 
$e^+ p \rightarrow e^+ X$ at $Q^2 > 15,000$ GeV$^2$ compared to Standard
Model (SM) expectations based on conventional
parton distribution functions (PDF).
There are several possibilities that might explain the observed anomaly 
at HERA\cite{1,2,qcd,3,4,5,6,7}.
First, a statistical fluctuation could be responsible although the
probability for this is small\cite{1}.
 A second possibility is that the PDF's
at high-$Q^2$ and large $x$ are  not well understood.
Some new effects may modify PDF's in the relevant 
kinematic region\cite{2,qcd}. And finally, there is the possibility 
that there are some new physics sources which cause the 
anomalous events.
Many theoretical explanations 
introducing new physics have been proposed\cite{3,4,5,6,7}. 
If data from HERA indeed represent a genuine 
deviation from the SM, it is 
important  to discover some way by which one can distinguish 
different explanations for the anomalous events 
using 
data from HERA. 

In all current known investigations, the basic strategy
in search for whatever new physics is to look for the 
direct excess from the 
prediction of the SM. 
This, however, sensitively depends on the statistics of the
experimental result and on how precisely 
the SM result is calculated. Unfortunately, the uncertainty of the 
calculation may be 
quite large due to our poor knowledge about the PDF's 
at high-$Q^2$ and large $x$. 
In this paper we will introduce a new measure 
which is defined as a ratio of 
the cross section integrated over different angular ranges. The advantage 
of doing this is that the uncertainty due to parton distributions would be 
remarkably reduced. 
This measure would make it possible to distinguish different 
models with less ambiguity as the statistics of data is
improved in the near future. 

The cross section for $e^+p \rightarrow e^+ X$ is given by
\begin{eqnarray}
{d\sigma(e^+p)\over dx dy}= {sx\over 16 \pi} \sum_q \{
q(x,Q^2)[|M_{LR}^{eq}|^2 +|M_{RL}^{eq}|^2 +(1-y)^2(|M_{LL}^{eq}|^2
+|M_{RR}^{eq}|^2)]\}\;,
\end{eqnarray}
where the summation on q is over all valence and sea quarks,
$Q$ is the momentum transfer, $x = Q^2/(2P\cdot k)$ and
$y = Q^2/sx$ are the Bjorken variables, $P$ is the proton momentum,
$k^2 = -Q^2$, $ s$ is the CM energy squared, 
$q(x,Q^2)$ is the q-quark PDF,
and 
\begin{eqnarray}
M_{ij}^{eq} &=& -{e^2Q_eQ_q\over sxy} -{g_Z^2(T^3_{ei} - sin^2\theta_W
Q_e)(T^3_{qj} - sin^2\theta_W Q_q)\over sxy+m_Z^2} + M_{ij}^{eq}(new)\;.
\end{eqnarray}
Here $Q_f$ and $T^3_{fi}$ are the charges and weak isospins of
relevant fermions, respectively,
$g_Z = e/(sin\theta_W cos\theta_W)$.
In the above equation the first two terms in $M_{ij}$ are contributions
from the SM and the last term is from possible new physics beyond the SM.

For concreteness, we will consider three representative possibilities
for the explanation of the anomaly:
1) Modification of PDF's; 2) New physics due to  s-channel production of 
new particle; 
And 3) New physics due to contact interactions.

If the anomaly is due to the modification of the PDF's, 
there is no change in parton level cross section.
There are strong constraints on the PDF's from experimental 
 data at low $Q^2$.
However, in the kinematic region relevant to HERA data the
PDF's are not well understood, especially for large $x$.
Modification required to the PDF's 
may arise 
from non-perturbative dynamics 
which produces new effects in the PDF's at low
$Q^2$ and very large $x \sim 1$. This new effect will migrate down to 
smaller $x$ as one flows up to higher $Q^2$ due to QCD evolution. 
Without the detailed understanding of the dynamics, one can not
calculate the new contributions, but can try to parameterize
possible new contributions and to fit available experimental data.
It has been shown in Ref.\cite{2} that by adding to the u-quark
PDF a small contribution at low $Q\sim 1.6$ GeV and $x> 0.75$,
one can significantly  enhance the event rate at large $Q^2$ and large $x$. 
In our later discussion, we will not attempt to fit the new contribution
to all other experimental data, instead we will concentrate on the
possibility of distinguishing this explanation from
others discussed below by  using HERA data.

In our numerical analysis
we will assume that the new effect produces a rectangular peak 
in the u-quark PDF at $Q \sim 1.6$ GeV 
and  $x >0.9$
(the specific shape of the new effect is not important). 
The size of the peak is determined
by fitting the $e^+ p \rightarrow e^+ X$ data from H1 and ZEUS. 
In our calculation, the CTEQ4M PDF's are used as conventional ones.
We find that the additional term with $0.5\sim 1\%$ (as mentioned in Ref.
\cite{2})
of the conventional valence u-quark PDF at $Q\sim 1.6$ GeV and $x > 0.9$ 
is not enough to produce the data. At least $4\%$ is 
required. The effects of modification of the PDF's are shown 
in Fig.1.

If the anomaly is due to s-channel resonance, a 
new particle with appropriate mass will be produced. 
The new particle may be a leptoquark or a stop with R-parity violation.
Assuming the new particle $\phi$ of mass $m_\phi$ to be 
a scalar with coupling to $e$ and $q$ given by
$L = \lambda \bar e_i q_j \phi$, the change in the cross section 
is to have a non-zero $M_{ij}^{eq}(new)$ equals to 
$ -|\lambda|^2/(2(xs-m_{\phi}^2 + i\Gamma m_{\phi}))$. 
Here $\Gamma $ is the total decay width of $\phi$.

H1 data indeed suggest a resonance with the excess
events clustered at $x$ between 0.4 and 0.5 which  translates into the 
determination of the mass about 200 GeV\cite{1}.
If the leptoquark is coupled to $e_L$ and $u_R$, to explain the 
anomalous events the parameter $\lambda$ is determined to be $0.025$\cite
{4,5}. 
The result is shown in Fig.1. 
 
If the anomaly is due to new contact interactions, which may arise from 
compositeness of quarks or exchange of heavy particles beyond the SM, the 
new effect will modify the parton level scattering matrix element. 
The relevant contact interaction is conventionally parameterized as\cite{6}
\begin{eqnarray}
L_{C.T.} &=& \sum_q [\eta_{LL}^{eq} 
(\bar e_L \gamma_\mu e_L \bar q _{L}
\gamma^\mu q_L) + \eta_{RR}^{eq}
( \bar e _R \gamma_\mu e_R \bar q_R \gamma^\mu q_R)
\nonumber\\
&+& \eta_{LR}^{eq}( \bar e_L\gamma_\mu e_L \bar q_R \gamma^\mu q_R) 
+\eta_{RL}^{eq}( \bar e_R \gamma_\mu e_R \bar q_L \gamma^\mu q_L)]\;.
\end{eqnarray}
The change in the cross section is to have  a non-zero
$M_{ij}^{eq}(new)$ equals to $\eta_{ij}^{eq}$.

There are many constraints on allowed values for parameters
$\eta_{ij}$ which limit possible solutions for the anomaly\cite{4,5,6,7}. 
Two types of solutions have been proposed in the literature. They are
solutions {\bf A}\cite{6} and {\bf B}\cite{7}:
\begin{eqnarray}
\eta_{LR}^{eu} = \eta_{RL}^{eu} = 1.4 \mbox{TeV}^{-2} \hspace{0.6in}\mbox{and} 
\hspace{0.6in}\eta_{RL}^{eu} = -\eta_{RR}^{eu} = 1.9 \mbox{TeV}^{-2},
\end{eqnarray}
respectively and all the other $\eta_{ij}$ equal to zero.
The value for solution {\bf B} is larger than that given  in Ref. \cite{7} because 
d-quark contribution was included according to gauge invariance there. 
The effects of these two solutions on the production rates are
shown in Fig.1 also.

From Fig.1 it is clear to see that
 all the models discussed above can equally explain 
the anomaly at HERA if only the production rate is concerned.
However it is 
not possible to distinguish these 
models from each other. 
In order to distinguish different models, more information is
needed. The obvious way to obtain such information is to study 
differential cross section. As have been mentioned 
previously that the leptoquark production 
model is featured by a resonance in the $x$ distribution,
whereas modification of PDF's and introduction of contact interactions would
not produce resonance structure. This can be used to discriminate 
the leptoquark production  from other models. 

To further distinguish other models, one must look for some 
 other distributions. 
 The differential distribution in $x$ and $y$
 may provide some needed information. In order to avoid 
the trouble from our poor understanding about the PDF's which strongly 
influences the shape of this distribution,  it would be important
to study distributions whose dependence on the PDF's is eliminated, 
or at least is substantially reduced. To this end we introduce  
a new measure 
to distinguish different models.

It is obvious that if one uses the  ratio of cross sections in two 
regions with 
given $x$ and  $y$, the dependence on PDF's will be greatly reduced. 
This will not entirely eliminate the PDF's effects because
the cross section is summed over different quarks inside the proton.
However, if one of the quark contribution dominates over others, the 
cancellation will be significant. This is true for the cases 
considered here, for example, the contribution from u-quark 
in the SM is more 
than 90\% for $Q^2> 15,000$ GeV$^2$. It, however, would 
require a tremendous luminosity 
to have a reasonable statistical significance,
if  the differential cross section in both $x$ and $y$ is used.
A traditional solution 
to this is to investigate the integrated cross section, 
say over the variable $y$. 
The integration procedure, however,  may weaken the cancellation of PDF's.
Therefore, the key to this investigation is to find a quantity
on which both the PDF's and the 
cross section do not change dramatically. If these conditions 
are satisfied, the PDF's can 
almost be factorized out from the integral of the cross section 
over such a quantity. Naively, from the first term 
in $M_{ij}$, one would think that the
dependence of the cross section 
 on $1/y$ is a constant for a fixed $x$ which may be good
 for our study. Numerically, we find $\psi = 1/y^2$ is more suitable
because of the appearance of the additional terms.  The cross section as a function of $\psi$ is shown in Fig.2 for fixed $x$.
 One finds indeed that 
the distribution is fairly flat. The new measure we suggest in this paper
is defined as
\begin{eqnarray}
R_\psi (x) = {\int^{\psi_{mid}}_{\psi_{low}}( d^2\sigma/dx d\psi)d\psi
\over \int^{\psi_{up}}_{\psi_{mid}} (d^2 \sigma/dx d\psi)d\psi}\;.
\label{rpsix}
\end{eqnarray}

We also note that in the region under consideration the PDF's very weakly
depend on the variable $\psi$.
This can be easily seen from Fig.2. 
 Therefore, the cancellation of PDF
effects does not sensitively depend on how the integration regions for 
numerator and denominator are divided in Eqn.\ref{rpsix}, although the value
of $R_{\psi}$ does. It is clear that 
$R_{\psi}(x)$ will be insensitive to the PDF's, but is sensitive to changes
in the parton level cross section and therefore to new physics.

Numerically, we used 100, 20 and 1 for $\psi_{up}\;, \psi_{mid}$, and 
$\psi_{low}$, respectively. 
As expected, for 
the SM the ratio is almost flat as $x$ varies from 0.2 to 1.
This is shown in Fig.3.
Results for different models are also calculated and plotted 
on the same figure. From the figure we see indeed that
changes in PDF's do not alter the distribution very much. 
As promised that the distribution is sensitive to new physics,
the s-channel new particle production and contact interaction 
models show very different features.

The leptoquark production model produces a distribution very close 
to the prediction of the SM in almost all the regions of $x$ except 
in the vincinity where $x$ is close to the resonance. A very high and 
very sharp peak is predicted corresponding to the very short life time 
of the possible resonance, whose width is about $\lambda^2 m_\phi /16\pi$, i.e. 
$\sim$ 6.3 MeV. 
The peak is caused by the interference between the real part of the 
resonance with the SM amplitude at close to the resonance 
($m_\phi^2 \pm \Gamma m_\phi$) 
and at the resonance by the imaginary part of the resonance amplitude.
The height of the peak depends on
how we divide the integration regions. 
The peak is a prominent feature for s-channel new 
particle production model. For easy comparison with experimental 
data, the effect of the leptoquark production
is plotted in a histogram 
instead of the peak in Fig.3. 
It is an effect averaged over the whole bin 
from 0.4 to 0.5 in which data provide a
reasonable statistical significance as described below.

The contact interaction model also has very distinctive feature
from the PDF modification model. The 
distributions for both solutions {\bf A} and {\bf B} mentioned previously
rise considerably at large $x$. This is a clear signature for contact 
interaction. 

In Fig.3, we also present our analysis of the available
data. With H1's 14 pb$^{-1}$ and ZEUS's 20 pb$^{-1}$ data\cite{1}, we have
443 events with $Q^2>2500$ GeV$^2$ and 191 events with  $Q^2>5000$ GeV$^2$. 
The point indicated by an open circle is obtained using H1 data only ($x<0.4$) because 
ZEUS data are not available in large $\psi$ region.
Above 0.4, data from both groups are  used which are indicated by filled 
circles. The luminosity 
normalization is entirely canceled since the ratio, $R_{\psi}$, is taken.
At large $x$ , the bin size on $x$ axis is taken quite big because
of the poor statistics. 
We also carried out an exercise assuming that data points in large $\psi$
and low $x$ region for ZEUS, 
where the data are not available, is similar to 
H1 data. The total event number in this region is then assumed to be
$(1+ L_{ZEUS}/L_{H1})$ times H1 event number. 
Here $L$ means luminosity. Now, there are much more data in the region 
with $x<0.4$, we divide this region into two bins.
These points are indicated by squares in Fig. 3.
The experimental results are drawn on top
of the theoretical predictions. Because of the big
error bars associated with the data points, it is 
hard to say which 
model is ruled out. 
The combined data points seem to favor 
the leptoquark production explanation.  However, 
we must keep in mind that the two points indicated by 
squares are not directly obtained from experimental data. 
The assumption we made may not be 
valid. We suggest our experimental colleagues to carry 
out a thorough analysis.

Before concluding our paper, we would like to comment on several
other tests for different models.
If the anomaly in $e^+p\rightarrow e^+ X$ is indeed due to
modification of the PDF's, there should be anomaly in $e^-p \rightarrow
e^- X$ also because the cross section in these two modes are scaled
approximately by the same amount. It will be tested once more 
data are available.
If the modification of the PDF is on the u-quark PDF, similar enhancement
will show up in $e^-p\rightarrow \nu X$, whereas if the modification is on
d-quark PDF, a large enhancement will show up in $e^+ p\rightarrow \bar \nu X$
which is in conflict with the data already\cite{8,9}.

The particular scalar s-channel
production model discussed before 
will only have anomaly in $e^+ p \rightarrow e^+X$\cite{4}.
Whereas for contact interaction model, it is possible to have
either result. Due to the particular chiral structure of solution {\bf A},
$e^- p \rightarrow e^-X$ is suppressed by a factor of $(1-y)^2$ and
has little effect on the 
SM cross section. However, solution {\bf B} will have
enhanced $e^-p\rightarrow e^- X$ due to large $\eta_{RR}^{eu}$.
In order to distinguish these models from the PDF modification model,
one needs to study the $R_\psi(x)$ distribution discussed before.

The predictions for charged current for s-channel 
resonance and contact interaction 
models are very model dependent\cite{10}. 
It is difficult to draw generic
conclusions without specific model which we will not attempt in this paper.

To conclude, we have introduced a new method to distinguish models for 
the high-$Q^2$ $e^+ p \rightarrow e^+ X$ anomaly at HERA. For concretness,
three representative models have been considered. 
Using this new method it is possible to distinguish different 
models with increased luminosity.
It is clear that the same analysis can be used to distinguish
other models. Also
the same analysis can be used for $e^- p\rightarrow e^- X$,
$e^-\rightarrow \nu X$ and
$e^+ p\rightarrow \bar \nu X$. We urge our experimental colleagues to carry 
out such analyses.

This work was supported in part by the US Department of Egergy under Grant
No. DE-FG03-96ER40972 and the
Australian Research Council.

\begin{figure}[htb]

\caption {The integrated cross section of $e^+p\rightarrow e^+ X$ 
for different models.
Experimental data are taken from Ref.[1].
The CTEQ4M PDF's are used as the conventional ones.}
\end{figure}

\begin{figure}
\caption{ The differential cross section of SM and sum of 
parton distribution at given $x$ versus $\psi$. 
For easy comparison 
with each other, the results for $x$=0.5 and 0.7 are multiplied 
by a factor of 10 and 100 respectively.}
\end{figure}

\begin{figure}
\caption{ 
The ratio $R_{\psi}$ versus $x$ for different models. 
Data are taken from Ref.[1].
A sharp peak 
for the leptoquark production is marked
by the arrow. The leptoquark contribution to the whole bin from 0.4 to 0.5 
is represented by the histogram and the rest of the 
ratio is exactly the same as the SM prediction.}

\end{figure}

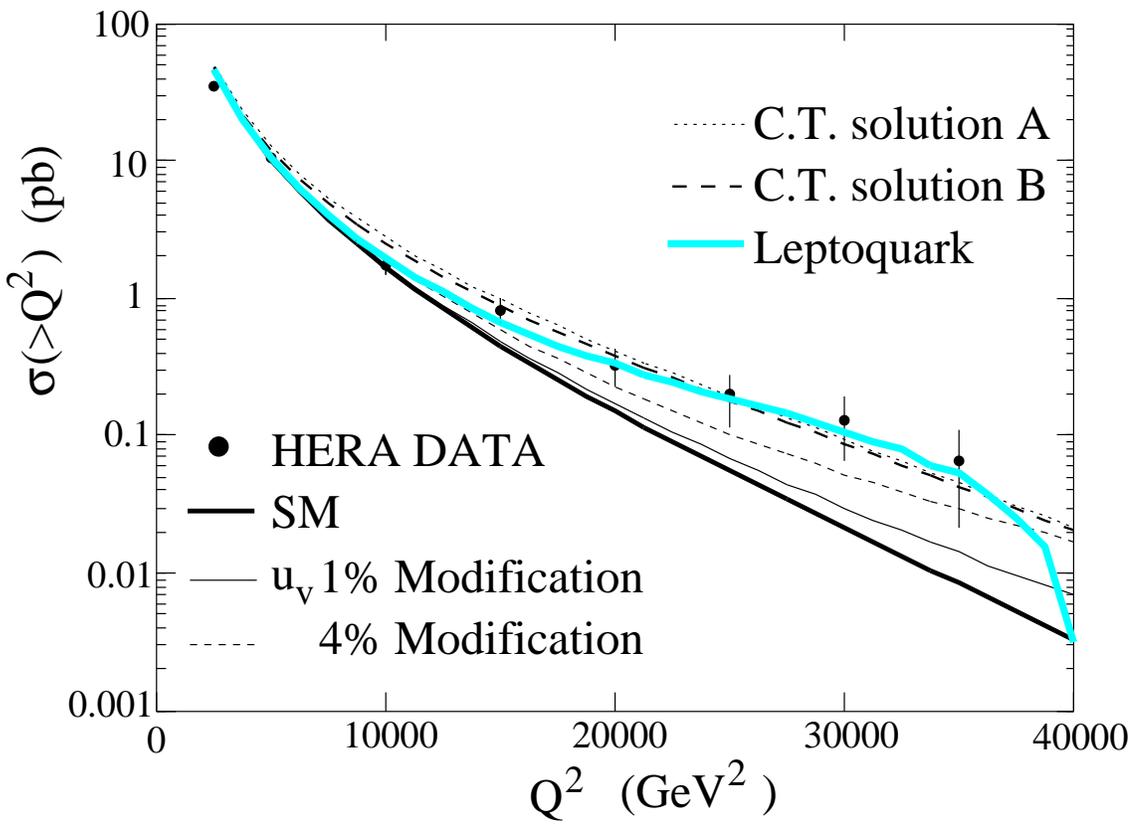
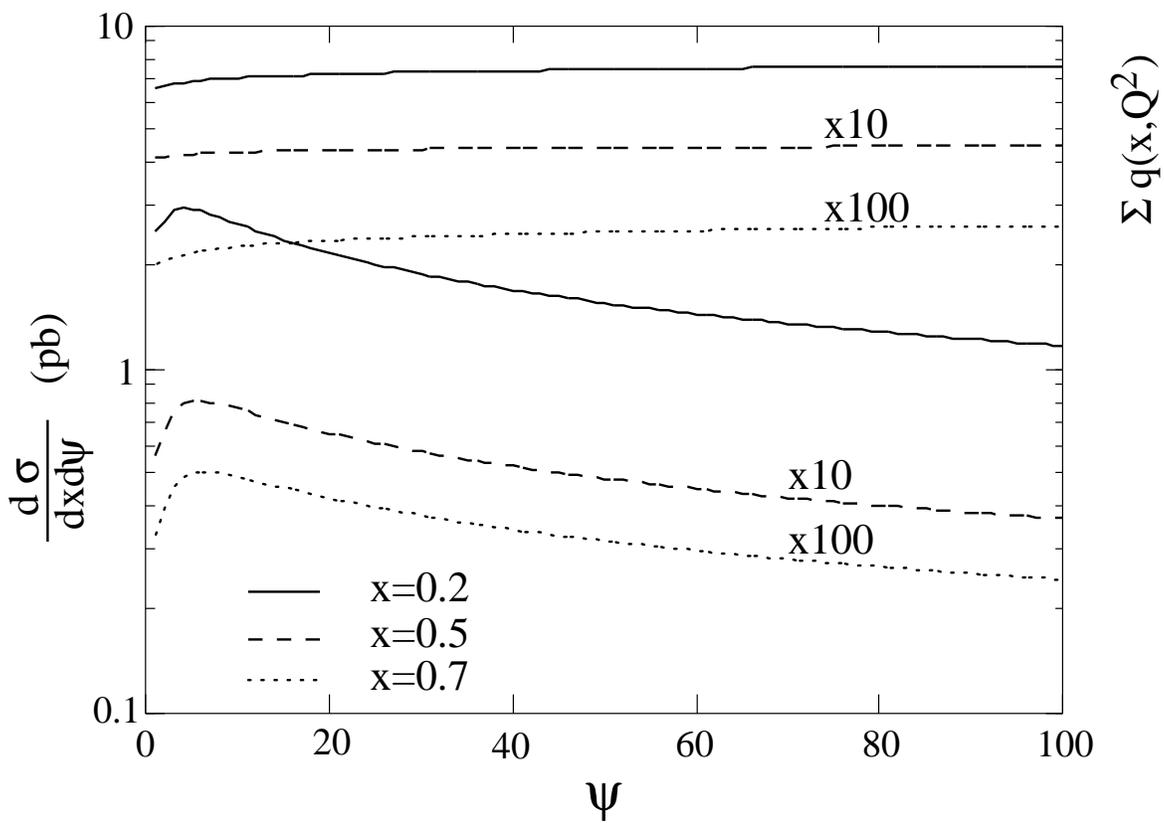
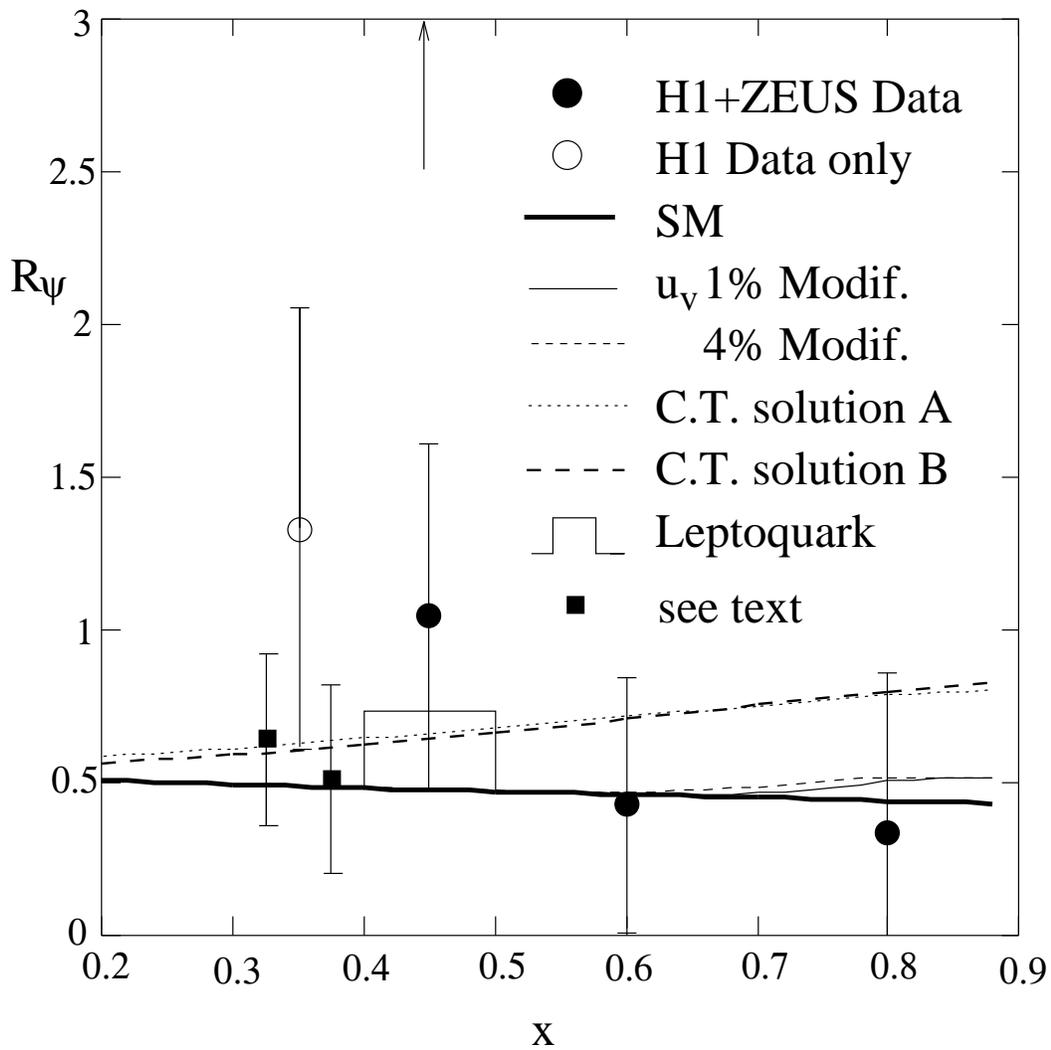
\end{document}